\documentclass[aps,twocolumn,superscriptaddress,reprint,notitlepage,longbibliography]{revtex4-1}
\usepackage{graphicx, color}
\usepackage{amsmath, amssymb, amsfonts, mathrsfs, ulem, comment}
\usepackage{times}
\usepackage[margin=1in]{geometry}

\begin{document}

\title{Molecular Dynamics Simulations on Cloud Computing and Machine Learning Platforms}

\author{Prateek Sharma}
\email{prateeks@iu.edu}
\affiliation{Intelligent Systems Engineering, 700 N. Woodlawn Avenue, Indiana University, Bloomington, Indiana 47408}

\author{Vikram Jadhao}
\email{vjadhao@iu.edu}
\affiliation{Intelligent Systems Engineering, 700 N. Woodlawn Avenue, Indiana University, Bloomington, Indiana 47408}

\begin{abstract}
Scientific computing applications have benefited greatly from high performance computing infrastructure such as supercomputers. 
However, we are seeing a paradigm shift in the computational structure, design, and requirements of these applications. 
Increasingly, data-driven and machine learning approaches are being used to support, speed-up, and enhance scientific computing applications, especially molecular dynamics simulations. 
Concurrently, cloud computing platforms are increasingly appealing for scientific computing, providing ``infinite'' computing powers, easier programming and deployment models, and access to computing accelerators such as TPUs (Tensor Processing Units). 
This confluence of machine learning (ML) and cloud computing represents exciting opportunities for cloud and systems researchers. 
ML-assisted molecular dynamics simulations are a new class of workload, and exhibit unique computational patterns. These simulations present new challenges for low-cost and high-performance execution.  
We argue that transient cloud resources, such as low-cost preemptible cloud VMs, can be a viable platform for this new workload. Finally, we present some low-hanging fruits and long-term challenges in cloud resource management, and the integration of molecular dynamics simulations into ML platforms (such as TensorFlow). 
\end{abstract}

\maketitle

\section{Introduction}

Scientific computing applications play an important role in the analysis and understanding of a wide variety of natural and synthetic processes. 
These applications are typically implemented as large-scale parallel programs that use communication frameworks such as MPI, and are largely deployed on high performance computing (HPC) infrastructure such as supercomputers. 
Molecular dynamics (MD) simulations are among the most ubiquitous scientific computing applications. These simulations have been used extensively by materials scientists, chemical engineers, and physicists to investigate the microscopic origins of the macroscopic behavior of materials such as self-assembled nanoparticles, viral capsids, electrolytes, lubricants, and polymers \cite{marson2015rational,hagan2016recent,ewen2018advances,anousheh2020ionic}.

A key goal for MD simulations is to explore the design space associated with the material attributes, and establish the links between the design parameters and the material response (generally, encoded in the structural and dynamical properties). 
To this end, simulations are deployed as a collection or ``bag'' of jobs. 
Collectively, a bag of jobs ``sweeps'' a multi-dimensional design space and furnishes the links between the material design parameters (inputs) and the material response (outputs). 
These links provide a reliable guide to experiments for a rational discovery of regions of the material design space exhibiting interesting structural and dynamical properties. 

The bags of jobs approach is also central to the rapidly developing area of using machine learning (ML) to enhance MD simulations and expedite the exploration of the material design space \cite{ferguson2017machine,wang2019machine,casalino2021ai,moradzadeh2019molecular,aspuru2019,sun2019deep,kadupitiya2020machine,kadupitiya2020machine2,kadupitiya2020simulating,kadupitiya2021probing}. 
Large collections of jobs with independent parameter sets are launched in order to train and test the ML models designed to enhance the predictive power or reduce the computational costs of MD simulations. 
For example, artificial neural network based regression models, trained on data from MD simulations of soft materials, can successfully predict the relationships between the input parameters and the simulation outcomes \cite{kadupitiya2019machine,kadupitiya2020machine2}. 
These ML surrogates accurately predicted the distributions of ions for a variety of confined electrolyte systems with $95\%$ accuracy and an inference time $10000\times$ less than the corresponding MD simulation runtime \cite{kadupitiya2019machine,kadupitiya2020machine2}.
As the utility of MD simulations and their ML-enhanced versions in the rational design of materials is further demonstrated, it will be necessary for accurate and fast simulations to be performed for larger sets of parameters in order to efficiently explore the material design space. To this end, it is important to leverage diverse advanced cyberinfrastructure platforms to perform low-cost simulations. 

\section{Molecular Dynamics on Cloud Platforms}

Increasingly, cloud computing platforms have begun to supplement and complement conventional HPC infrastructure to meet the large computing and storage requirements of simulations \cite{buyya-hpc-survey}. Public clouds offer many benefits: on-demand resource allocation, convenient pay-as-you-go pricing models, and ease of deployment on an ``infinite'' resource pool. An important objective in cloud deployments is to optimize for cost in addition to performance. Costs can be reduced through the use of transient computing resources that can be unilaterally revoked and preempted by the cloud provider, but their preemptible nature results in frequent job failures. 
The considerations of cost, frequent job failures, and server configuration heterogeneity intrinsic to the system present multiple challenges in deploying applications on cloud platforms. These challenges are fundamentally different from those that appear in using HPC clusters as the execution environment for simulations.

Systems such as  SciSpot~\cite{kadupitige2020modeling,scispot-github}, a framework that uses a new reliability model for constrained preemptions of Google Preemptible Virtual Machines (VMs), can optimize the deployment of scientific computing applications on transient cloud servers and enable low-cost MD simulations.
SciSpot uses an empirical and analytical model of transient server availability to predict expected running times and costs associated with jobs of different types and durations.
Considering an entire bag of jobs as an execution unit enables simple and powerful policies for optimizing cost, makespan, and ease of deployment. 
SciSpot's cost-minimizing server selection and job scheduling policies reduce costs by up to $5\times$ compared to conventional cloud deployments. 

\section{Molecular Dynamics on ML Systems}

The use of ML systems such as TensorFlow and PyTorch has been largely limited for designing ML-based enhancements (e.g., surrogates, integrators, force fields) for MD simulations \cite{kadupitiya2020machine2,wang2019machine,kadupitiya2020simulating}. Some recent studies have explored the utilization of ML platforms for ``non-ML'' tasks related to MD simulations \cite{yao2018tensormol,schoenholz2020jax,gao2020torchani,torchmd,barrett2020hoomd}. 
However, the use of ML systems to develop and execute MD simulations and integrate them with ML-based enhancements is unexplored.
In the following, we outline advantages of ML systems as environments for executing MD simulations and integrating them with data-driven models. We also discuss the associated systems challenges. 

MD simulations are typically coded in C/C++ and parallelized using OpenMP and MPI \cite{lammps.plimpton}. However, more modern MD software packages such as HOOMD-Blue \cite{anderson2020hoomd} have demonstrated that MD simulations can be easily written in high-level languages such as Python. ML platforms such as TensorFlow are based on Python and the associated high-level data-flow abstraction can enable rapid prototyping of MD simulations. We highlight a few key advantages of utilizing ML  systems for executing MD simulations:

\begin{itemize}
    \item High-performance simulations: ML systems offer automatic parallelization of simulations and enable seamless use of next-generation cloud and HPC hardware such as GPUs and TPUs. \item One-stop platform: ML systems offer extensive support for debugging, data analytics, and post-processing tasks that can be leveraged to perform all simulation-related tasks at a single platform.
    \item Large ecosystem: Developers have access to a much bigger data science and ML community. 
    \item Better software engineering: Associated tools are richer and are actively developed and improved (compared to, for example, MPI).
    \item Reproducibility and ease of sharing: Users and developers can easily share all simulation models and methods in one notebook that can be run on ML system backends such as Google Colab (compared to cumbersome configuring of simulations on different HPC systems).
    \item Integration with data-driven approaches: Data operations in ML systems are  first-class operations, instead of being a separate stage of the simulation workflow aimed at exploring the material design space.
\end{itemize}

The seamless integration of simulations and data-driven models on a single platform offers many opportunities for a rational and expedited exploration of the material design space. We now illustrate a set of these opportunities using ML surrogates as an example. An ML surrogate is a model trained on data from MD simulations that is used to approximate the relationships between the input parameters and the simulation outcomes, bypassing part or all of the explicit evolution of the simulated components. 
For instance, the ML surrogate in~\cite{kadupitiya2020machine2} can reduce the inference time from 30 minutes to 0.2 seconds---a $10,000\times$ speedup! 
This surrogate model was trained using a bag of jobs of size $N=6,000$,  
each job representing a unique set of parameters discretizing the high-dimensional input design space. 

In the conventional, un-integrated approach, the bag of jobs (of size $N$) is run sequentially on HPC systems, with a long wait time until a surrogate is trained. 
$N$ is chosen \textit{a priori} (usually informed by domain expertise) such that the generated data is sufficient to train an ML model (typically a deep neural network).
In contrast, an integrated approach utilizing ML systems for simulation facilitates a rapid transition from simulation to surrogate. 
The surrogate training can now begin with a smaller number $n$ of simulations and the training progress can be monitored in a seamless fashion. When surrogates are designed in an integrated manner on ML systems, a number of new opportunities emerge:

\begin{itemize}
    \item A unified approach to executing simulations and training ML models to approximate input-output relationships enables the development of the surrogate during the exploration of the material design space with a bag of $n < N$ jobs. ML systems can facilitate the automation of the switch to deriving outputs using surrogates when the training and testing errors become small and surrogate accuracy reaches a high value. 
    \item On-the-fly development of the surrogates enables a principled approach to quantify the completion of the material design space exploration.
    \item Surrogate accuracy and inferences times can be readily improved with minimal overhead associated with the one-stop platform that enables a seamless accumulation of training data resulting from more simulation executions.
    \item The ease of designing surrogates in parallel with the simulation-driven exploration of the material design space enables efficient new simulation code development (e.g., writing new pair interaction potentials). The trained surrogates can be used as benchmarks of existing understanding that can guide code updates. 
\end{itemize}

\section{Open Questions and Future Directions}

The confluence of ML, MD simulations, and cloud computing presents the broader cloud and HPC research communities with several exciting challenges, and also provides a unique opportunity to bring these communities together. 
ML-assisted MD simulations impose a unique set of computational requirements which will require advances in cloud resource allocation, such as: 
\begin{itemize}
    \item \textbf{Abstractions:} Our bags-of-jobs abstraction is the first step towards a ``cloud-native'' abstraction for ML-assisted MD workloads. Easier ways to deploy such applications on the cloud will lower costs and will make it easier for domain scientists to rapidly iterate. 
    \item \textbf{Rethinking performance metrics:} Conventional metrics such as parallel speedups will be insufficient for ML-assisted workloads that use a combination of training and inference, where the answer can be provided by a trained ML-model.
    We believe that cost will remain a first-level metric in the cloud, and must be a core part of performance and resource optimizations. 
\end{itemize}

We also claim that ML platforms 
can provide a single, unified framework for future applications that will integrate ML and MD simulations in new ways. Fundamentally, we are proposing to use systems in ways that they are not designed for, which leads to many natural performance challenges. 
For instance, the dataflow model used by TensorFlow provides suboptimal performance for fine-grained parallelism required for MD simulations on GPU and TPU clusters, and new performance optimizations and abstractions are necessary. 

Finally, bringing a ``classic'' HPC workload such as MD simulations into an entirely new cloud+ML ecosystem will require the HPC, cloud, and ML communities to work together with domain scientists and engineers in new ways. 
The confluence will provide opportunities for the next generation of domain scientists to be trained in practical ML and cloud skills. At the same time, cloud researchers should be more cognizant of this new class of workload, which is radically different from the ``enterprise'' and ``big-data'' workloads that cloud platforms have traditionally been optimized for. 

\section*{Acknowledgments}
\vspace{-5pt}
{V.J. was partially supported by NSF through Award 1720625.}

\end{document}